\documentclass[PRB,twocolumn]{revtex4}
\usepackage{}
\usepackage{epsfig}
\usepackage{graphicx}
\usepackage{amsfonts,amssymb}
\usepackage{amsmath}
\usepackage{color}
\usepackage {times}
\definecolor{refcolor}{rgb}{1.0,0.0,0.0}
\usepackage{lipsum}
\usepackage[colorlinks,citecolor=blue,linkcolor=blue,anchorcolor=blue,filecolor=blue,urlcolor=blue]{hyperref}

\newcommand{\be}{\begin{equation}}
\newcommand{\ee}{\end{equation}}   
\newcommand{\bea}{\begin{eqnarray}}
\newcommand{\eea}{\end{eqnarray}}
\newcommand{\ba}{\begin{array}}
\newcommand{\ea}{\end{array}}

\newcommand{\q}{{\bf q}}
\renewcommand{\k}{{\bf k}}

\begin{document}

\title{Non-reciprocal spin-wave excitations in Rashba-Hubbard ferromagnets}

\author{Aastha Jain and Dheeraj Kumar Singh}
 
\affiliation{Department of Physics and Material Science,
Thapar Institute of Engineering and Technology, Patiala 147004, Punjab, India}

\date{\today}


\date{\today}
\begin{abstract}
We explore the nonreciprocity of spin-wave excitations in the Rashba-Hubbard ferromagnet on  a square lattice. Our study reveals that the propagation of spin-wave excitations exhibit non-reciprocal behavior, i.e., spin waves traveling in opposite directions display asymmetry in energy dispersion $\omega({\bf q}) \ne \omega(-{\bf q)}$, which  also results in an  asymmetric behavior of group velocity, spin stiffness, etc. We find that this asymmetric behavior arises only when the magnetic moments are aligned inside the atomic plane, while the excitations remain symmetric for out-of-plane magnetization. The first dominating term in the low-energy dispersion is linear. However, if the magnetic moments are out-of-plane, then the first dominant term is quadratic instead. The low-energy non-quadratic behavior is examined in the intermediate-to-strong coupling regime for various strengths of Rashba spin-orbit coupling.
\end{abstract}

\maketitle

\section{Introduction}
 Magnonics and spintronics have attracted considerable interest in recent times because of their huge potential technological applications~\cite{kru,lenk,chumak,barman,wolf,zu}. Magnonics and spintronic devices are often based on systems such as crystals without inversion symmetry or interfaces of heterostructures, which culminates in locking of spin and momentum, i.e., the Rashba spin-orbit coupling (SOC)~\cite{rashba,chumak,manchon,dyak}. The Rashba SOC generates the Dzyaloshinskii-Moriya type interactions (DMI)~\cite{D,M}, which is known to compete with the symmetric Heisenberg-exchange interaction~\cite{bode}, leading to a plethora of unusual ordering phenomena such as chiral magnetism, spatially extended magnetic orders~\cite{D2,D3,M2,miya,togawa}, noncollinear spin textures including magnetic skyrmions, etc~\cite{uk,muhl,yu,heinze}. Moreover, material systems with active SOC and inversion symmetry can also exhibit interesting but complex phases and associated low-energy excitations~\cite{mohapatra1,mohapatra2}.
 
  DMI can also induce the magnon Hall effect when a thermal gradient drives a spin wave current~\cite{onose}. In ferromagnets, the domain walls can acquire specific chirality, allowing them to be moved efficiently by applied currents~\cite{ryu,emori}. These domain walls are considered fundamental components for magnonic logic~\cite{jamali,garcia,garcia2}. In antiferromagnets, DMI causes domain walls to function as a spin-wave polarizer following Malus's law, as a spin-wave retarder at high frequencies~\cite{lan}, and as a transistor~\cite{cheng,kosty}. In other words, the SOC-induced DMI can potentially alter the spin-wave dynamics in magnetic materials~\cite{ma,lan2, udvardi,costa,moon,cort,santos2}.
  
  Given the known heat losses and size limitations encountered with semiconductor devices, new information processing technologies are being intensively explored~\cite{markov,kaji}. These technologies do not rely solely on charge as the carrier of information but instead exploit the spin, intrinsic degree of freedom of electrons. In light of this, spin waves are emerging as a promising candidate in the field of spintronics and magnonics, offering an alternative to the conventional complementary metal oxide-semiconductor devices~\cite{kru,lenk,chumak,barman,dieny}. They represent the collective excitations of electronic spins in magnetically ordered systems and can offer significant advantages~\cite{bloch}. They propagate through a material without causing any heat losses, thereby enabling energy-efficient transport of spin current. Second, they have relatively shorter wavelengths, which makes them valuable for further miniaturizing devices~\cite{neusser}.

The spin-wave excitations can be probed using inelastic neutron scattering (INS)~\cite{pet}, resonant inelastic x-ray scattering (RIXS)~\cite{poel,kim}, and spin-polarized electron energy loss spectroscopy  (SPEELS)~\cite{pli,vollmer}. In recent years, considerable experimental efforts have been devoted to the experimental investigation of the nonreciprocal nature of spin-wave excitations in various systems. Zakeri et al., studied  double Fe layers on W (110) by using SPEELS and demonstrated an asymmetry in the spin-wave dispersion~\cite{Zakeri}. Subsequent studies further revealed the role of spin-orbit coupling (SOC) in the directional dependence of magnon propagation~\cite{zakeri1}. Employing SPEELS, the asymmetry in magnon dispersion relation of Co/Fe bilayers grown on W (110) was studied. The energy asymmetry of magnon mode has been measured and compared with Fe bilayers grown on W (110)~\cite{tsurk}. Evidence of nonreciprocity has also been reported in Pt/CoFeB ultrathin films~\cite{kai} and Pt/Co/AlO$_x$ films using Brillouin light scattering (BLS)~\cite{bel}. Furthermore, heterostructures such as W/CoFeB/SiO$_2$~\cite{ak} and NiFe/CoFeB~\cite{heins} bilayer systems with varying thicknesses display nonreciprocal spin-wave propagation.

 As demonstrated mostly through the localized spin models with DMI in the presence or absence of symmetric Heisenberg interaction, a term linear in the wave vector for the spin-wave dispersion lifts the chiral degeneracy of magnons and leads to an asymmetric dispersion~\cite{costa,moon,santos2}. In other words, the spin-wave excitations with wavevectors of equal magnitude but opposite directions are no longer degenerate. This nonreciprocal behavior arises only when the magnetization and DMI vectors are not perpendicular. By measuring this asymmetry, the strength and chirality of the DMI can be determined.
 
 Despite intensive research in this direction~\cite{udvardi,costa,moon,santos2,cort,kostyl}, the spin-wave excitations in magnetically ordered system with Rashba SOC has been paid less or almost no attention, particularly, in the microscopic model such as the  Rashba-Hubbard model. In such a model, the non-reciprocal behavior of the spin-wave excitations can provide important insight into the nature and strength of SOC as demonstrated through the current work. It is this aspect of the problem, which is the focus of the current work. Here, we investigate the non-reciprocal nature of spin-wave excitations in the one-orbital Rashba-Hubbard ferromagnet.
 
 The initial step in this direction is to look for a region in the parameter space where the ferromagnetic (FM) state may be stable. Most of the earlier studies exploring the phase diagram of the Rashba-Hubbard model have focused instead on the role of interplay between the SOC and on-site Coulomb interaction at half-filling~\cite{kennedy,kubo,ajain,ggoyal,hotta}. At this filling and depending on the SOC strength, a variety of phases including Weyl semimetal without N\'{e}el order, Weyl semimetal with N\'{e}el order, normal N\'{e}el order, spiral order, vortex states, etc. have been proposed to exist~\cite{ajain,kubo,sebas}. On the other hand, only a few studies have examined the phase diagram away from half filling~\cite{sudhakar,greco}. These investigations  explored the existence and stability of FM state as a function of band filling, strength of on-site Coulomb interaction, and strength of SOC parameter. For instance, one work provides the range of these parameters based on the study of static spin susceptibility~\cite{greco}. We choose various parameters accordingly to study the spin-wave excitations.

This paper is organized as follows. Section II describes the one-orbital Rashba-Hubbard model, mean-field decoupling of the on-site Coulomb interaction term, and the self-consistent scheme to obtain the ferromagnetic state. The method of obtaining the spin-wave excitation in the ferromagnetic state is also explained in this section. The results are presented in section III. Section IV concludes the manuscript with a brief discussion.

\section{Model and Method}
We consider the one-orbital Rashba-Hubbard Hamiltonian defined on a square lattice, which is given by
\begin{eqnarray} 
    \mathcal{H} &=& -\sum_{<{\bf i},{\bf j}>}\sum_\sigma ( t_{{\bf i}{\bf j}} d_{{\bf i}\sigma}^{\dagger}  d^{}_{{\bf j} \sigma} + h.c.) - \mu \sum_{i,\sigma} n_{{\bf i}\sigma}   \nonumber\\
  &+& \lambda \sum_{{\bf i},\sigma,\sigma^\prime} [ \textit{i}(d^\dagger_{{\bf i}\sigma}\sigma_{\sigma\sigma^\prime}^x d^{}_{{\bf
   i}+\hat{y},\sigma^\prime} - d^\dagger_{{\bf i}\sigma}\sigma_{\sigma\sigma^\prime}^y d^{}_{{\bf i}+\hat{x},\sigma^\prime})+ h.c.] \nonumber\\
   &+&  U \sum_{\bf i} \hat{n}_{{\bf i}\uparrow} \hat{n}_{ {\bf i} \downarrow}.
\end{eqnarray}  \label{1}
The first term denotes the kinetic energy, accounting for the energy gain due to the delocalization of electrons. Both nearest-neighbor and next-nearest-neighbor hopping are incorporated. $d_{{\bf i} \sigma}^{\dagger}$ ($d_{{\bf i} \sigma}$) is the creation (annihilation) operator for an electron at site ${\bf i}$ with spin $\sigma$. The second term represents the chemical potential. The third one describes the Rashba SOC or spin-flip process, with $\lambda$ denoting the SOC strength, and $\sigma^{i}$ standing for the Pauli matrices. The last term represents the on-site Coulomb repulsion between the electrons of opposite spins, where $\hat{n}_{{\bf i} \sigma} = d^\dagger_{{\bf i} \sigma}d^{}_{{\bf i} \sigma}$ is the number operator.

After Fourier transformation, the non-interacting part of the Hamiltonian becomes
\begin{eqnarray}
     H({\bf k}) &=& \sum_{{\bf k},\sigma} (\epsilon_{\bf k}-\mu) d_{{\bf k}\sigma}^\dagger d^{}_{{\bf k} \sigma} \nonumber \\
  &+ & \lambda \sum_{{\bf k},\sigma,\sigma'}d_{{\bf k} \sigma}^\dagger[2 \sin k_y \sigma_{\sigma \sigma^\prime}^x
  -  2 \sin k_x \sigma_{\sigma \sigma^\prime}^y]  d_{{\bf k} \sigma'}
\end{eqnarray}
with $\epsilon_{\bf k} = -2t(\cos k_x + \cos k_y) + 4t' \cos k_x \cos k_y$. Here, $t$ and $t'$ represent nearest and next-nearest neighbor hopping amplitudes. Throughout,  $t$ is used as a unit of energy.

\subsection{Hartree-Fock mean-field theory}
The Hubbard interaction term, being quartic in terms of electron-field operators, has been treated via variety of techniques based on the mean-field theoretic approaches, perturbation techniques~\cite{sene}, dynamical mean-field theory (DMFT)~\cite{georges}, classical and quantum monte-carlo~\cite{vekic}, variational monte carlo (VMC)~\cite{kubo2} etc. mostly in the absence of Rashba SOC. Here, we employ static mean-field approach based on the Hartree-Fock approximation to decouple the interaction term as our focus is mainly on the low-energy collective excitations at low temperature. The bilinear term in the electron-field operator, thus, obtained is
\begin{equation}
{H}_{im} 
=  {- \frac{U}{2} \sum_{{\bf i} \sigma} \psi^{\dagger}_{\bf i}(\bf { \sigma}\cdot{\bf m}_{\bf i}) \psi^{}_{\bf i}},
\vspace{-3mm}
\end{equation}
where $\psi^{\dagger}_{\bf i} = (d^{\dagger}_{{\bf i} \uparrow}, d^{\dagger}_{{\bf i} \downarrow}) $. The $j$-th component of magnetic moment at the site ${\bf i}$ is ${m}^{k}_{\bf i} = \frac{1}{2} \langle \psi^{\dagger}_{\bf i}{\sigma}^{k} \psi^{}_{\bf i} \rangle$ with ${\bf m}_{\bf i}$ is the magnetic moment. ${\sigma}^{k}$ is $k$-th Pauli matrix. Upon including the decoupled interaction term, the momentum-space Hamiltonian for the Rashba-Hubbard ferromagnet is
\begin{eqnarray}
   \mathcal{H}_{HF} &=& \sum_{{\bf k},\sigma} (\epsilon_{\bf k}-\mu ) d_{{\bf k}\sigma}^\dagger d^{}_{{\bf k} \sigma} +  \lambda \sum_{{\bf k},\sigma,\sigma'}d_{{\bf k} \sigma}^\dagger[2 \sin k_y \sigma_{\sigma \sigma^\prime}^x \nonumber\\
  &-&  2 \sin k_x \sigma_{\sigma \sigma^\prime}^y]  d_{{\bf k} \sigma'} - \sum_{{\bf k},\sigma,\sigma'}d_{{\bf k} \sigma}^\dagger ({\vec{\sigma}}_{\sigma \sigma^\prime} \cdot {\bf \Delta})d_{{\bf k} \sigma'}.
\end{eqnarray}
In the matrix form, the same can be expressed as
\begin{eqnarray}
\mathcal{H}_{HF} &=& \sum_{\mathbf{k}} \psi^\dagger_{\mathbf{k}\sigma} \left[ (\epsilon_{\mathbf{k}} -\mu) \sigma^{0} + (\lambda \, \vec{\alpha}(\mathbf{k}) - \vec{\Delta}) \cdot \vec{\sigma} \right] \psi_{\mathbf{k} \sigma} \nonumber\\
&=& \sum_{\mathbf{k}} \psi^\dagger_{\mathbf{k}\sigma} H_{fm} ({\bf k} ) \psi_{\mathbf{k} \sigma}, \nonumber\\
\end{eqnarray}
where $\sigma^{0}$ is a 2$\times$2 identity matrix,
$\vec{\alpha}(\mathbf{k}) = \left( 2 \sin k_y,\, -2 \sin k_x,\, 0 \right)$, and $\psi^\dagger_{\mathbf{k}\sigma} = (d^{\dagger}_{{\bf k}\uparrow}, d^{\dagger}_{{\bf k}\downarrow})$. The eigenvalues for the above Hamiltonian can be expressed as
\begin{eqnarray}
    \varepsilon_{1/2}(\k)& =&\epsilon_\mathbf{k} - \mu \pm \sqrt{V_{\k}},
\end{eqnarray}
where 
\begin{eqnarray}
 &V_{\k}=&4\lambda^2(\sin^2 k_x + \sin^2 k_y)
   - U \lambda(m_x \sin k_y - m_y \sin k_x )\nonumber\\
   &+& \frac{U^2}{4}(m_x^2 +m_y^2 +m_z^2).
\end{eqnarray}
\noindent The magnetic exchange-field is
\begin{equation}
\vec{\Delta} = \frac{U}{2} \vec{m},
\end{equation}
which is dependent on the magnetic moments given by
 \begin{eqnarray} 
m_z &=&\sum_{{\bf k},l}(\phi^{*}_{ {\bf k}\uparrow l} \phi^{}_{{\bf k}\uparrow l} - \phi^{*}_{ {\bf k}\downarrow l} \phi^{}_{{\bf k}\downarrow l}) f(E_{\k l})  \nonumber\\
m_x &=& \sum_{{\bf k},l}(\phi^{*}_{{\bf k}\uparrow l} \phi^{}_{{\bf k}\downarrow l} + \phi^{}_{{\bf k}\uparrow l} \phi^{*}_{{\bf k}\downarrow l})  f(E_{\k l}) \nonumber\\
m_y &=& \sum_{{\bf k},l}(-i\phi^{*}_{ {\bf k}\uparrow l} \phi^{}_{{\bf k} \downarrow l} + i \phi^{*}_{ {\bf k}\downarrow l} \phi^{}_{{\bf k} \uparrow l}) f(E_{\k l}).
\end{eqnarray}
$l$ is the band index and $f(E)$ is the Fermi-Dirac distribution. $(\phi_{\mathbf{k} \uparrow l}, \phi_{\mathbf{k} \downarrow l})$ is the eigenvector of the Hamiltonian matrix given by Eq. (5), which corresponds to the eigenvalue $E_{l}$. The magnetic moments in the FM state are determined in a self-consistent manner using Eqs. (5), (6), and (7).

\subsection{Methodology}
In order to investigate the spin-wave excitations in the FM state, we study the dynamical spin-susceptibility within the random-phase approximation (RPA)~\cite{greco} given by
\begin{equation} \label{eq_RPA_suscep}
\hat{\chi} ({\bf q},i\omega_l)
= {\hat{\chi}^{0} ({\bf q},i\omega_l)}{\left[ I -   \hat{U}\hat{\chi}^{0} ( {\bf q}, i \omega_l )    \right]}^{-1},
\end{equation}
where the bare susceptibility matrix $\hat{\chi}^{(0)}$ is
\begin{equation}
\hat{\chi}^{0}({\bf q,i\omega_l})= 
\left(
\begin{array}{llll}
\chi^{0}_{\uparrow \uparrow \uparrow \uparrow} & \chi^{0}_{\uparrow \downarrow \uparrow \uparrow} & \chi^{0}_{\uparrow \uparrow \downarrow \uparrow} 
&\chi^{0}_{\uparrow \downarrow \downarrow \uparrow} \\
\chi^{0}_{\uparrow \uparrow \uparrow \downarrow} & \chi^{0}_{\uparrow \downarrow \uparrow \downarrow} & \chi^{0}_{\uparrow \uparrow \downarrow \downarrow} 
&\chi^{0}_{\uparrow \downarrow \downarrow \downarrow} \\
\chi^{0}_{\downarrow \uparrow \uparrow \uparrow} & \chi^{0}_{\downarrow \downarrow \uparrow \uparrow} & \chi^{0}_{\downarrow \uparrow \downarrow \uparrow} 
&\chi^{0}_{\downarrow \downarrow \downarrow \uparrow} \\
\chi^{0}_{\downarrow \uparrow \uparrow \downarrow} & \chi^{0}_{\downarrow \downarrow \uparrow \downarrow} & \chi^{0}_{\downarrow \uparrow \downarrow \downarrow} 
&\chi^{0}_{\downarrow \downarrow \downarrow \downarrow}
\end{array}
\right) \; 
\end{equation} 
The elements of the bare susceptibility matrix $\chi^{0}_{\sigma_1 \sigma_2 \sigma_3 \sigma_4} ({\bf q},i\omega_l)$ is defined as the convolution of two Green's functions
\begin{eqnarray}
&&
\chi^{0}_{\sigma_1 \sigma_2 \sigma_3 \sigma_4} ({\bf q},i\omega_l)
=
\nonumber\\
&& \qquad
-\sum_{ {\bf k}, i\nu_n} G^{(0)}_{\sigma_1 \sigma_4}  ({\bf k},i\nu_n)
G^{(0)}_{\sigma_3 \sigma_2}  ( {\bf k}+ {\bf q},i\nu_n+i\omega_l),  \qquad
\label{chi0}
\end{eqnarray}
\noindent where
\begin{equation}
G^{0} ( {\bf k}, i\nu_n ) =
\left[i\nu_n \sigma_0- H_{mf} ({\bf k} ) \right]^{-1}
\label{2x2Green},
\end{equation}
$\k$ is a momentum in the Brillouin zone and $\nu_n$ is the Matsubara frequency.
The interaction matrix $\hat{U}$ is a $4 \times 4$ matrix with the form
\begin{equation}
\hat{U}= 
\left(
\begin{array}{llll}
0&0&0&U\\
0&0&-U&0\\
0&-U&0&0\\
U&0&0&0
\end{array}
\right) \;.
\end{equation}
It may be noted that one can easily obtain various components of the physical spin suscetibility such as $\chi_{xx}$, $\chi_{yy}$, etc.,  i.e., $\chi_{ij}(\q,i\omega_l)$ in terms of $\chi^{(0)}_{\sigma_1 \sigma_2 \sigma_3 \sigma_4} ({\bf q},i\omega_l)$ as follows
\be
\chi_{ij}(\q,i\omega_l) = \frac{1}{4}\sum_{\sigma_1, \sigma_2, \sigma_3,\sigma_4} \sigma^{i}_{\sigma_1, \sigma_2} \sigma^{j}_{\sigma_3, \sigma_4} \chi^{(0)}_{\sigma_1 \sigma_2 \sigma_3 \sigma_4} ({\bf q},i\omega_l).
\ee
The spin-wave excitations correspond to the pole of RPA susceptibility defined by Eq. 10. Therefore, it can be recognized as a peak of -Im$({\rm Tr}\hat{\chi} ({\bf q},i\omega_l))$ = -Im$\sum_{i} \hat{\chi}_{ii} ({\bf q},i\omega_l))$ when the latter is plotted as a function of frequency along the high-symmetry direction via analytic continuation $i\omega_l \rightarrow \omega + i \eta$. We chose $\eta$ = 0.001 for finite broadening.

\section{Results}
\subsection{Stability of Ferromagnetic Order}

\begin{figure}[h]
    \centering
    \includegraphics[scale=1, width= 8.8cm]{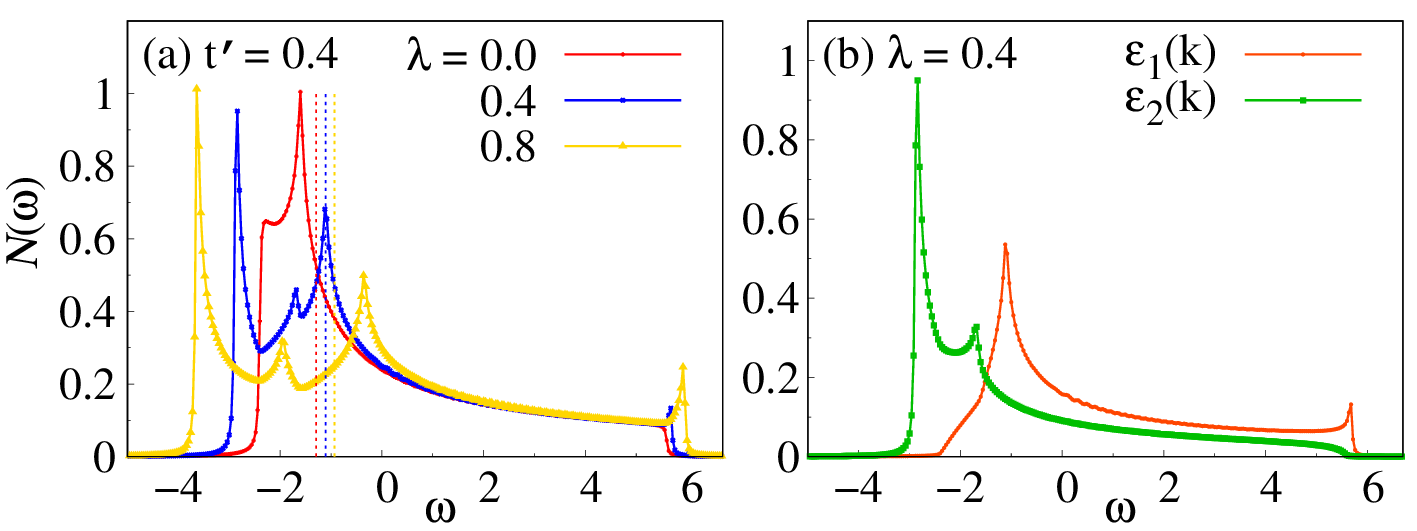}
    \caption{(a) The DOS in the unordered state for different SOC couplings. (b) For $\lambda = 0.4$, contributions from different band to the DOS are plotted.}
    \label{1}
\end{figure}

The majority of earlier works have suggested the stabilization of FM state with asymmetric density of states (DOS), which can result from a variety of factors, prominent ones being the lattice structure, long-range hopping, etc ~\cite{fark,fark2,fark3,sudhakar,spandey}. The studies, especially in two-dimensional systems, find that the FM state is stabilized when the Fermi level is not far away from the van-Hove singularity. The latter is known to be enhanced by the next-nearest hopping in a square lattice, besides introducing the asymmetry in the DOS~\cite{sudhakar,fark3}. In this direction, as it may be noted, earlier works were focused largely on the Hubbard model without RSOC. One recent study probing the role of RSOC, however, indicates that ferromagnetic fluctuations dominate within a specific region of the parameter space spanned by band filling, next-nearest neighbor hopping, and Rashba SOC. In particular, these ferromagnetic fluctuations are strong when the Fermi energy is not far away from the two van-Hove fillings or even in between them~\cite{greco}. The two van-Hove singularities result from the splitting of bands because of the RSOC, and the role of interband excitations is stressed in the stabilization of FM state.
\begin{figure}[h]
    \centering
    \includegraphics[scale=1.05, width= 8.8cm]{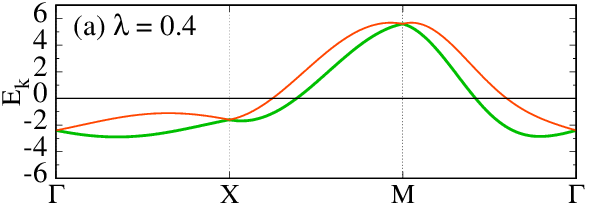}
    \includegraphics[scale=0.8, width= 8.8cm]{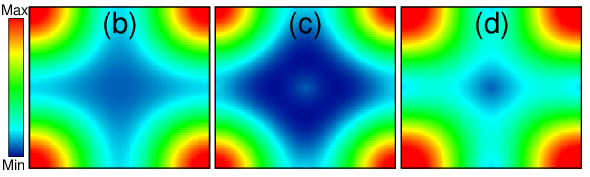}
    \caption{For the unordered state, (a) the electronic dispersion is shown for $\lambda = 0.4$ along the high-symmetry direction. (b) The electronic dispersion in the entire Brillouin zone for $\lambda = 0.0$. For $\lambda = 0.4$, the two split bands $\varepsilon_1(\k)$ and $\varepsilon_2(\k)$ are shown (c) and (d). }
    \label{2}
\end{figure}

\begin{figure}[h]
   \includegraphics[scale=1, width= 9.0cm]{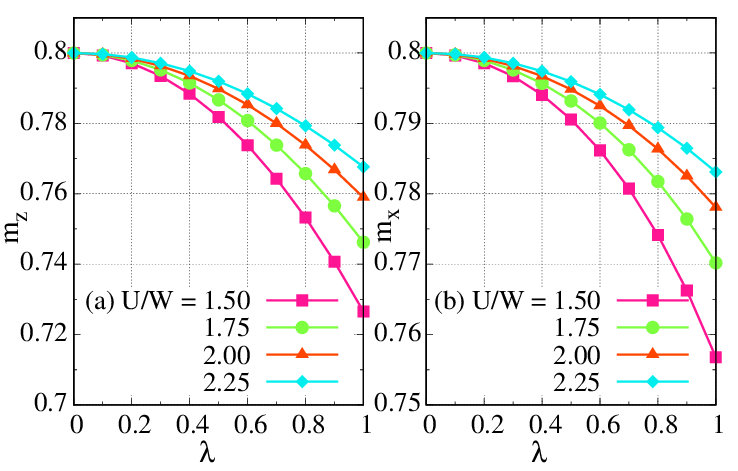}
    \caption{ (a) Out-of-plane magnetization $m_z$ and (b) in-plane magnetization $m_x$ as a function of RSOC for different strengths of $U$.}
    \label{3}
\end{figure}
Our study of spin-wave excitations begins with a set of parameters, where the FM state can be stabilized, as indicated by earlier studies. In the following, we consider the next-nearest neighbor hopping parameter $t' = 0.4$ and the electron filling $n = 0.8$ unless stated otherwise. For this set of parameters, the DOSs in the unordered state for different RSOC strengths are shown in Fig.~\ref{1}.

The electronic band dispersions are shown in the Fig.~\ref{2}(a). These results reveal the extent of band splitting or interband separation along the high-symmetry directions, when the RSOC is included. The two-fold degeneracy is removed except at $\Gamma$, X, and M.
For $\lambda = 0.4$, the Fermi energy is almost coinciding with one of the two van-Hove fillings resulting from RSOC splitting (see Fig.~\ref{2}(b) and (c)). It may be noted that there is an additional singularity towards the lower-band edge, which is the strongest of all. Interestingly, it does not correspond to a conventional saddle point, and rather results from the creation of a local maximum at $\Gamma$ in one of the dispersions and a moat-like dip running around it (Fig.~\ref{2}(c)). It may be interesting to see how this strongest band-edge singularity may affect the stability of FM state, which has been ignored in the earlier work.

As $\lambda$ varies within the range $0 \lesssim \lambda \lesssim 1$, the bandwidth $W$ varies in the range $ 8 \lesssim W \lesssim 9$. Bandwise, $W$ has different ranges for the first band $\varepsilon_1(\k) ( 8 \lesssim W \lesssim 8.5) $ and  second band $\varepsilon_2(\k) (8 \lesssim W \lesssim 9.5 )$. In order to express $U$ in terms of the bandwidth, we choose $W = 8$ throughout for simplicity .

Fig.~\ref{3} shows the self-consistently obtained magnetic moments for the range $ 0 \le \lambda \le 1$ in the two cases, i.e., when the magnetic moments are aligned in-plane and out-of-plane. With a rise in $\lambda$, the in-plane magnetic moment $m_x$ and out-of-plane magnetic moment $m_z$, both decrease, however, slowly. The decrease is slightly faster for $m_z$. We consider all types of orientation of magnetic moments in order to explore various possible interesting features of the spin-wave excitations.

\subsection{Spin Excitations in Rashba-Hubbard Ferromagnets}
In a ferromagnet, the time-reversal symmetry is intrinsically broken. In addition, the presence of the RSOC breaks the inversion symmetry, which can potentially affect the nature of the spin-wave dynamics in a significant way, as earlier studies in the localized-spin models indicate that the DMI can be responsible for a non-reciprocal behavior, i.e., $\omega({\bf q}) \ne \omega(-{\bf q)}$.

\begin{figure*}
    \centering
    \includegraphics[scale=1, width= 18cm]{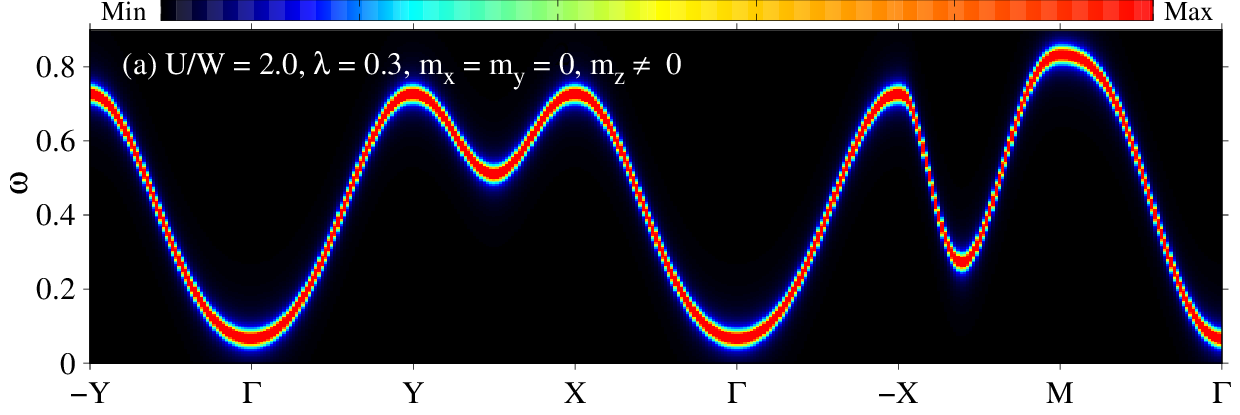}
    \includegraphics[scale=1, width= 18cm]{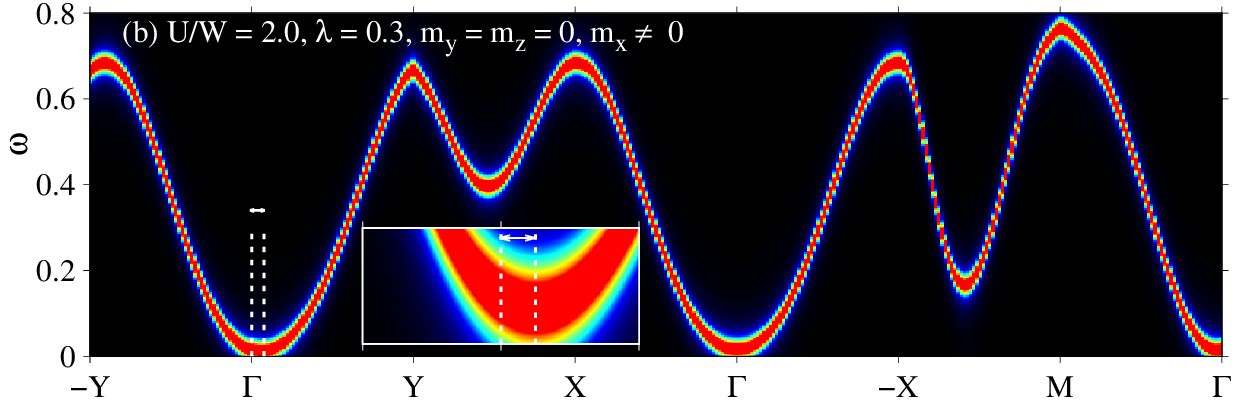}
    \includegraphics[scale=1, width= 18cm]{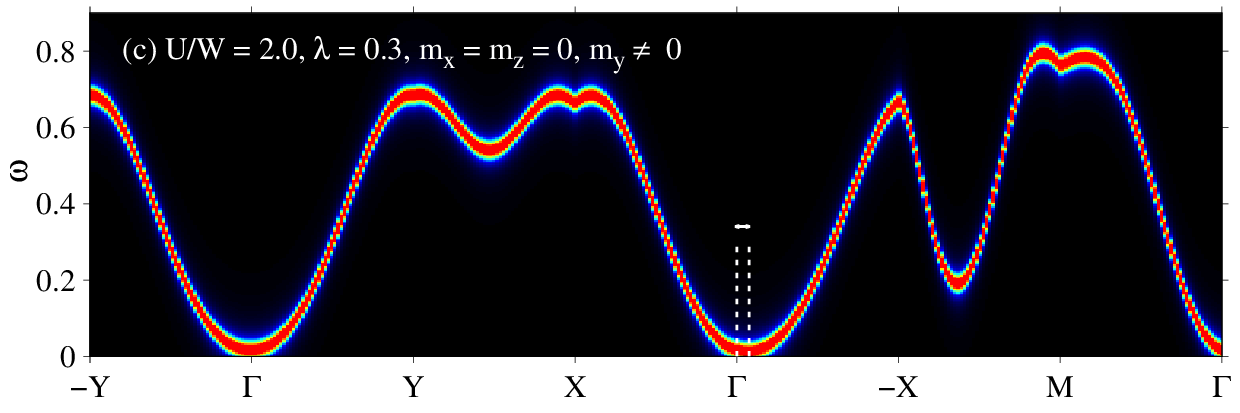}
    \caption{For $U = 2W$ and $\lambda = 0.3$, the spin-wave dispersion along the high-symmetry directions when the magnetic moments are aligned (a) out-of-plane and (b) in-plane ($x$-direction). Nonreciprocity is observed along -Y $\xrightarrow{}\Gamma\xrightarrow{}$ Y. In inset, a zoomed-in view of the shifted minima is shown, where the arrow indicates the shift. Similarly, in (c) moments are aligned in $y$-direction, and the non reciprocity is observed along X $\xrightarrow{}\Gamma\xrightarrow{}$ -X direction.}
    \label{4}
\end{figure*}
Our calculation finds that the spin-wave excitations in the Rashba-Hubbard ferromagnet exhibit a non-reciprocal behavior as seen in the Fig.~\ref{4}. The latter shows the spin-wave excitations for different orientations of magnetic moments, where $U/W = 2.0$ and $\lambda = 0.3$. The results clearly demonstrate that reciprocal behavior ($\omega({\bf q}) = \omega(-{\bf q)}$) exists only when the magnetization is aligned out-of-plane, whereas the non-reciprocal behavior ($\omega({\bf q}) \ne \omega(-{\bf q)}$) is observed for in-plane magnetization. In the latter case, the minimum of spin-wave excitation energy shifts away from $\Gamma$ point and there is a cuspy dip/rise near X, Y, and M points as well. In particular, an important difference between the spin-wave excitations for the  magnetizations along $x$- and $y$-directions is that there is a cuspy rise for the former while a cuspy dip for the latter. A more detailed analysis of these results is provided below.

\subsection{{\bf Out-of-plane Magnetic Moment}}

\begin{figure}[h]
    \centering
   \includegraphics[scale=1, width= 9.0cm]{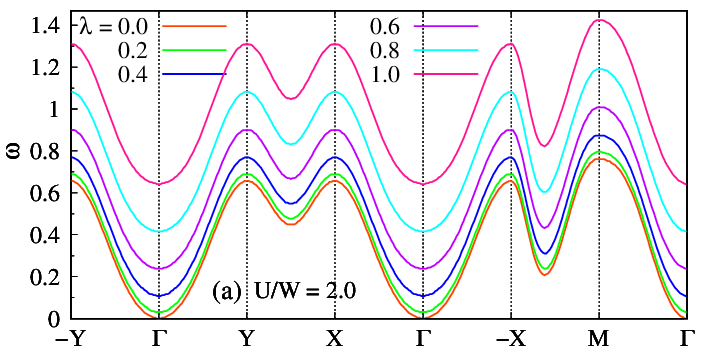}
    \includegraphics[scale=1, width= 9.0cm]{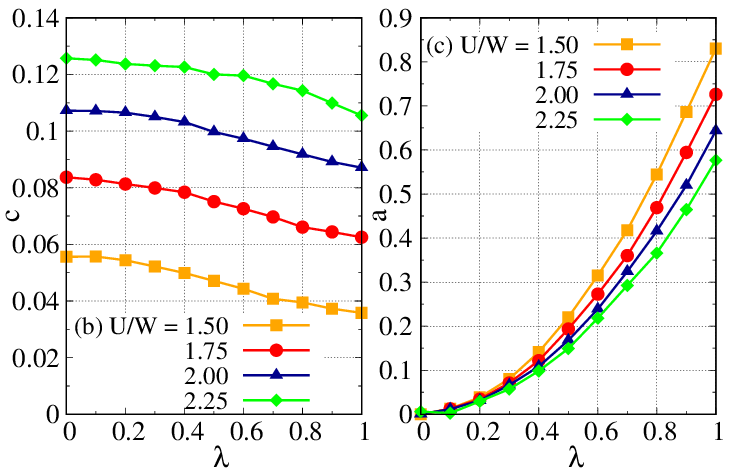}
    \caption{For out-of-plane magnetic moments, (a) the spin-wave dispersion for different RSOC strength, (b) coefficient of $q^2_{x/y}$, and (c) the constant term, as a function of RSOC and $U$.}
    \label{5}
\end{figure}

First, we examine the spin-wave excitations when the magnetic moments are aligned out of atomic plane, i.e., along the $z$-axis. Fig.~\ref{5}(a) shows the spin-wave dispersion along the high-symmetry directions, which exhibits symmetric behavior of the spin waves for the opposite momenta around $\Gamma$. The symmetry indicates the absence of non-reciprocal spin-wave excitations.

In the absence of RSOC, the Hamiltonian is invariant under spin rotations, exhibiting $SU(2)$ symmetry. It leads to a zero-energy mode at the $\Gamma$-point, i.e., $\q = (0, 0)$, in accordance with the Goldstone’s theorem. However, once the RSOC is introduced, the rotational symmetry is broken, and a gap opens at $\Gamma$. The gap increases significantly with the RSOC strength $\lambda$ (Fig.~\ref{5}(c)). The low-energy behavior of the dispersion follows $\omega_{q_{x/y}} \sim a + cq_{x/y}^2$, and there is no linear term. Both the gap $a$ and stiffness constant $c$ show dependence on $\lambda$ and $U$ (Fig.~\ref{5}(c) and (b)). As expected, $c$ increases and decreases with a rise in $U$ and $\lambda$, respectively. An increase in the spin-wave stiffness $c$ with $U$ suggests enhanced hardening of the FM phase. In contrast, the stiffness decreases with increasing RSOC strength, indicating a softness of the FM order, as illustrated in Fig.~\ref{5}(b). On the other hand, the effect of RSOC on the gap $a$ is more pronounced, which is reflected in its sharp rise with $\lambda$, whereas $U$ dependence is relatively weaker ( Fig.~\ref{5}(c)).

\subsection{ {\bf In-plane Magnetic Moment}}
If magnetic moments point along the $x$-direction, the spin-wave excitations exhibit asymmetry, i.e., $\omega_{\bf \q} \ne \omega_{\bf -\q}$  along the -Y $\xrightarrow{}\Gamma\xrightarrow{}$ Y. However, they remain reciprocal along the orthogonal direction -X $\xrightarrow{}\Gamma\xrightarrow{}$ X.  When the magnetic moments are aligned along the $y$-direction, a similar behavior is observed. In this case, the nonreciprocity appears along the X $\xrightarrow{}\Gamma\xrightarrow{}$-X direction but not along Y $\xrightarrow{}\Gamma\xrightarrow{}$ -Y direction.
This directional nonreciprocity or asymmetry clearly indicates the presence of an effective anti-symmetric DMI, which can arise due to the breaking of inversion symmetry by the RSOC.

\begin{figure}[h]
    \centering
    \includegraphics[scale=1, width= 8.8cm]{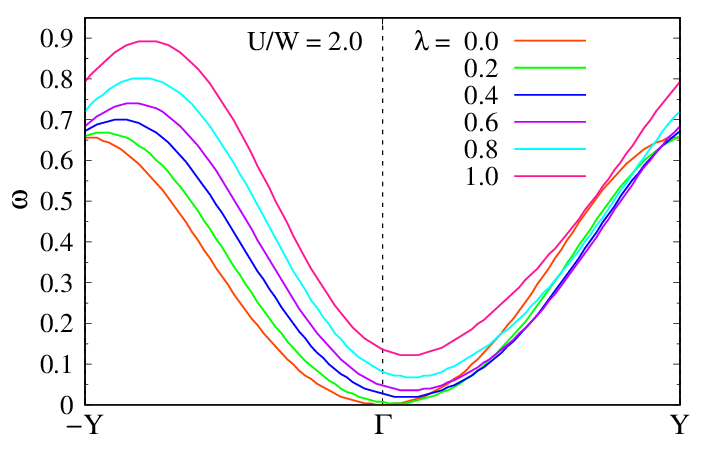}
    \caption{For the magnetic moments along $x$, the non-reciprocal behaviour of spin-wave dispersion is observed along -Y$\rightarrow \Gamma \rightarrow$Y.}
    \label{6}
\end{figure}

The non-reciprocal spin-wave dispersion is accompanied with a shift of minima away from $\Gamma$ together with a gap opening at the $\Gamma$-point. Both effects are clearly visible in the Fig.~\ref{6}, when the magnetic moments are aligned along $x$-direction. For clarity, the spin-wave dispersion is plotted only along -Y $\xrightarrow{}\Gamma\xrightarrow{}$ Y for $U = 2W$ and different strengths of the RSOC.

\begin{figure*}[t]
    \centering
    \includegraphics[scale=1.4, width= 16cm]{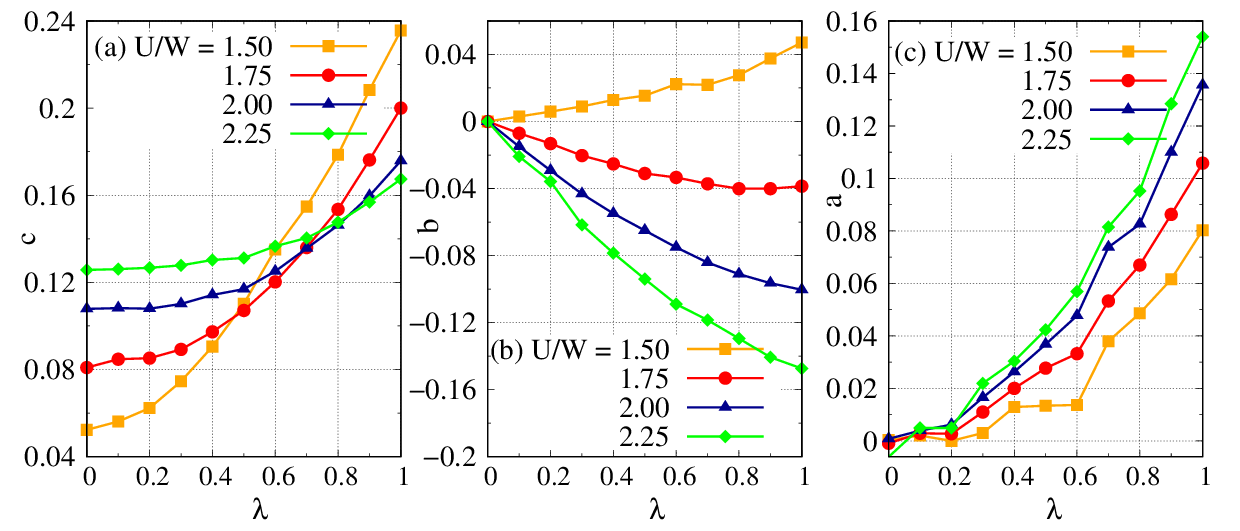}
    \caption{When the magnetic moments are pointing along the $x$-direction, behavior of the coefficients of (a) $q^2_{y}$ and (b) $q_{y}$, and (c) the constant term as a function of RSOC for different correlation strengths.}
    \label{7}
\end{figure*}

The shift of minima in the spin-wave dispersion near $\Gamma$ is indicative of the presence of a linear term in the low-energy behavior when modeled by $\omega_{q_{x/y}} \sim a+bq_{x/y}+cq^2_{x/y}$. The detailed analysis of the low-energy dispersion, shown in Fig.~\ref{7}, reveals that the linear term is significant, and it increases with RSOC. It is also dependent on the correlation strength $U$. For $U = 1.50W$, the dispersion minimum shifts to the left, indicating a positive coefficient ($b$) of the linear term. However, as $U$ increases to $1.75W$ or beyond, the sign of $b$ reverses, leading to a rightward shift of the dispersion minimum. For a fixed $\lambda$, the magnitude of this shift increases with increasing $U$. Thus, the nonreciprocity of the dispersion arises from a subtle interplay between both RSOC and electron-electron interaction.

In contrast to the case of magnetic moments along $z$ direction, here, the spin stiffness $c$ rises, as the RSOC strength increases. This effect is more pronounced at lower values of $U$, suggesting that in systems with weaker electron-electron interactions, the energy cost of spin twisting is very sensitive to RSOC. Moreover, there is a region near $\lambda \sim 0.6$, where $c$ may exhibit a weak dependence on $U$. However, moving away from this point, the dependence on $U$ increases.

\begin{figure}[!h]
    \centering
   \includegraphics[scale=1, width= 9.0cm]{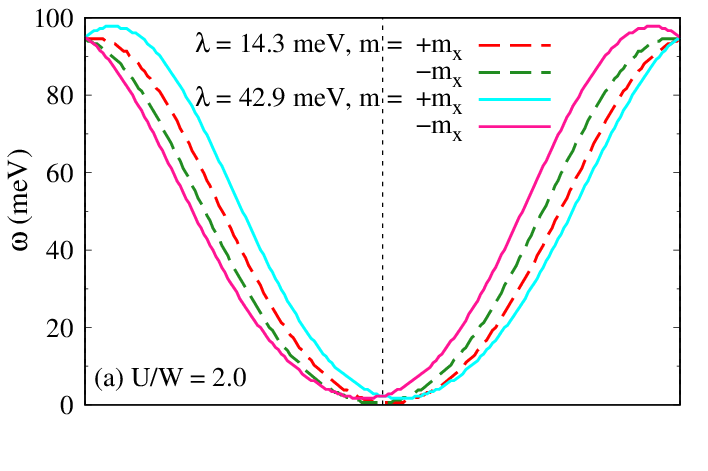}\\[-20pt]
    \includegraphics[scale=1, width= 9.0cm]{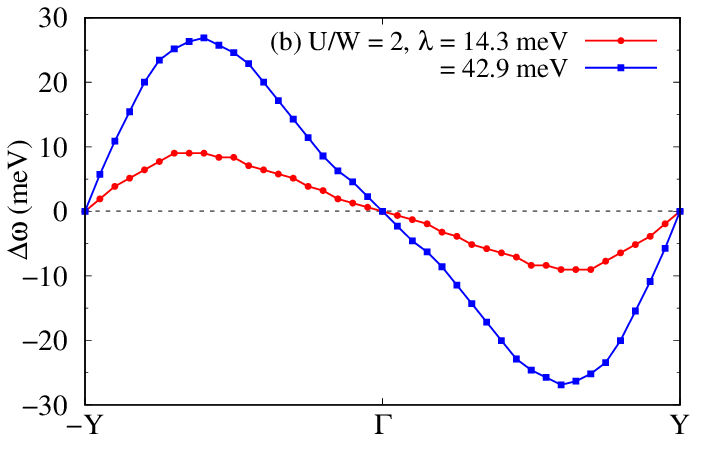}
    \caption{In (a) spin wave dispersions for two different RSOC strengths with opposite magnetization are plotted and (b) shows the difference between energies for spin wave excitations with same momentum but with an opposite magnetization direction for the two RSOC strengths. The hopping parameter is chosen to be $t = 143$~meV .}
    \label{8}
\end{figure} 
 
Moreover, the constant term $a$ in the low-energy dispersion, corresponding to the gap near dispersion minimum, also shows an overall increase with both the RSOC strength and the correlation strength. Its behavior is similar to the case when the magnetic moments are pointing out of the plane, however, the magnitude is significantly smaller. Thus, these combined effects highlight once again the crucial role of the interplay between direction of magnetic moments, RSOC, and electron correlations in shaping the spin-wave characteristics in the Rashba-Hubbard ferromagnets.
\section{Summary and Conclusions}

Spin-wave excitations, in the system without inversion symmetry, have been studied earlier, particularly due to their potential applications in spintronics and magnonics devices~\cite{chumak,garcia,garcia2,cheng,kosty}. However, majority of them used localized-spin models with DM interactions~\cite{udvardi,moon,santos2,cort,Zakeri}. Second, another set of work have investigated the effect of $LS$ coupling on the spin-wave excitations in iron thin films by employing multiorbital correlated models, especially to reproduce the experimental spin-wave characteristics, where multiple SOC parameters as well as on-site repulsive Coulombic parameters belonging to both inter-orbital and intra-orbital SOCs were used~\cite{abate,costa}. However, the study did not describe how the interplay between the spin-orbit coupling associated with the absence of inversion symmetry and repulsive Coulomb interaction can result in nonreciprocity.

Through the current work, we aimed to fill this gap and provide a detailed study of spin-wave excitations in the FM state of correlated systems with active Rashba SOC, using a microscopic and one-orbital Rashba-Hubbard model. Our work establishes the nonreciprocal nature of the spin-wave excitations in the Rashba-Hubbard ferromagnets. The spin-wave dispersion is strongly dependent on both the orientation of the magnetic moments and the direction of propagation. The nonreciprocity occurs when the magnetic moments are aligned in the plane, which is a result of linear term in the low-energy dispersion introduced by a subtle interplay of RSOC and on-site Coulomb interaction. The observed nonreciprocity for in-plane magnetization and specific wave vector directions indicate the presence of an effective in-plane antisymmetric DM interaction in the system.

Most of the experimental works have investigated the heterostructures and
ultrathin films with non-square interfacial lattice
structures. Therefore, a direct comparison of our results obtained for a square lattice with the experiments is not possible currently. In these systems, nonreciprocity typically may arise from the SOC, thickness dependent asymmetry in the presence of magnetic field, etc. Another experimentally observed feature is that the spin waves with the same momentum but opposite magnetization directions exhibit an asymmetry, i.e., for a given momentum $\q$, $\omega(m_x) \neq \omega(-m_x)$ as shown in Fig.~\ref{8}(a). The energy difference arising from this asymmetry upon reversing the direction of magnetization can be seen in  Fig.~\ref{8}(b). These results are qualitatively similar to the ones observed  experimentally for a ferromagnetic Fe double layer on W(110) with bcc-like structure by Zakeri \textit{et al.}~\cite{zakeri1}, demonstrating the nonreciprocal nature of spin waves upon magnetization reversal. Thus, it can be inferred that the presence of RSOC introduces a nondegeneracy in the energy of spin waves with opposite magnetization directions, and this asymmetry increases with increasing RSOC strength.

As the previous studies in the localized models suggest, the nonreciprocity in the spin-wave propagation may arise when the condition ${\bf m} \cdot {\bf D}_{ij} \neq 0$ is fulfilled, where ${\bf D}_{ij}$ is the DM vector and ${\bf m}$ is the magnetization direction~\cite{udvardi,santos2}. The observed left-right asymmetry in the spin-wave propagations suggests that the induced DM vector must lie in the plane perpendicular to the aligned magnetic moments.
We have analyzed the low-energy spin-wave dispersion by fitting it to a second-order polynomial form. This examining of the dispersion allows us to gain a qualitative understanding of other properties of spin waves, such as group velocity, spin stiffness, etc. Due to the asymmetry in the dispersion, spin waves propagate with unequal group velocities $({\bf v}_g)$ in opposite directions, i.e., ${\bf v}_g(\q) \neq {\bf v}_g(-\q)$, where ${\bf v}_g = {\nabla_{\q} \omega}$. In the small-$\q$ limit, the linear coefficient $b$ in the dispersion corresponds to the group velocity, which increases with RSOC strength as well as exhibits dependence on the on-site Coulomb interaction. Another important property, the spin stiffness, also increases with increasing SOC coupling and correlation strength, reflecting enhanced rigidity of the spin system against long-wavelength fluctuations.

To conclude, we have examined the spin-wave excitations in the Rashba-Hubbard ferromagnet on a square lattice. In addition to being gapped near the high-symmetry point $\Gamma$, the spin-wave excitations exhibit non-reciprocal behavior when the magnetic moments are aligned in the atomic plane. The non-reciprocal behavior is marked by a linear term in the low-energy dispersion, which is also responsible for the shift of dispersion minimum away from $\Gamma$. The linear term depends on both the Rashba spin-orbit coupling as well as on-site Coulomb interaction, indicating it to be a result of the interplay between the two factors. Thus, the current work provides important insights into the nature of spin-wave excitations at the interfaces with square-lattice arrangement without inversion symmetry via a microscopic Rashba-Hubbard model, which can be helpful in understanding the microscopic origin and degree of asymmetry in terms of spin-orbit coupling and correlation strength in future experiments.


\begin{thebibliography}{100}

\bibitem{kru} V. V. Kruglyak, S. O. Demokritov, and D. Grundler, Magnonics, \href{https://doi.org/10.1088/0022-3727/43/26/264001}{J. Phys. D: Appl. Phys. \textbf{43}, 264001 (2010).}

\bibitem{lenk} B. Lenk, H. Ulrichs, F. Garbs, and M. Münzenberg, The building blocks of magnonics, \href{https://doi.org/10.1016/j.physrep.2011.06.003}{Phys. Rep. \textbf{507}, 107 (2011).}

\bibitem{chumak} A. V. Chumak, V. I. Vasyuchka, A. A. Serga, and B. Hillebrands, Magnon spintronics, \href{https://doi.org/10.1038/nphys3347}{Nat. Phys. \textbf{11}, 453 (2015).}

\bibitem{barman} A. Barman, G. Gubbiotti, S. Ladak, A. O. Adeyeye, M. Krawczyk, J. Grafe, C. Adelmann, S. Cotofana, A. Naeemi, V. I. Vasyuchka, 
The 2021 magnonics roadmap, \href{https://doi.org/10.1088/1361-648X/abec1a}{J. Phys. Condens. Matter \textbf{33}, 413001 (2021).}

\bibitem{wolf}S. A. Wolf, D. D. Awschalom, R. A. Buhrman, J. M. Daughton, S. von Molnar, M. L. Roukes, A. Y. Chtchelkanova, and D. M. Treger, Spintronics: a spin-based electronics vision for the future, \href{https://doi.org/10.1126/science.1065389}{Science \textbf{294}, 1488 (2001).}

\bibitem{zu} I. Žutić, J. Fabian, and S. Das Sarma, Spintronics: fundamentals and applications, \href{https://doi.org/10.1103/RevModPhys.76.323}{Rev. Mod. Phys. \textbf{76}, 323 (2004).}



\bibitem{rashba}Y. A. Bychkov and E. I. Rashba, Properties of a 2D electron gas with lifted spectral degeneracy, \href{http://jetpletters.ru/ps/1264/article_19121.shtml}{JETP Letters \textbf{39}, 66 (1984).}

\bibitem{manchon} A. Manchon, H. C. Koo, J. Nitta, S. M. Frolov, and R. A. Duine, New perspectives for Rashba spin-orbit coupling, \href{https://doi.org/10.1038/nmat4360}{Nat. Mater. {\bf 14}, 871 (2015).}

\bibitem{dyak} M. I. Dyakonov and V. I. Perel, Possibility of orienting electron spins with current, 
\href{http://jetpletters.ru/ps/1587/article_24366.pdf}{Zh. Eksp. Teor. Fiz. \textbf{13}, 657 (1971)}.


\bibitem{D} I. Dzyaloshinsky, A thermodynamic theory of “weak” ferro-
magnetism of antiferromagnetics, \href{https://doi.org/10.1016/0022-3697(58)90076-3}{J. Phys. Chem. Solids {\textbf 4}, 241 (1958).}

\bibitem{M} T. Moriya, Anisotropic superexchange interaction and weak ferromagnetism, \href{https://doi.org/10.1103/PhysRev.120.91}{Phys. Rev. \textbf{120}, 91 (1960).}

\bibitem{bode} M. Bode, M. Heide, K. von Bergmann, P. Ferriani, S. Heinze, G. Bihlmayer, A. Kubetzka, O. Pietzsch, S. Blügel, and R. Wiesendanger, Chiral magnetic order at surfaces driven by inversion asymmetry, \href{https://doi.org/10.1038/nature05802}{Nature \textbf{447}, 190 (2007).}


\bibitem{D2} I. E. Dzyaloshinskii, Theory of helicoidal structures in antiferromagnets.I.nonmetals, \href{http://jetp.ras.ru/cgi-bin/dn/e_019_04_0960.pdf}{Sov. Phys. JETP \textbf{19}, 960 (1964).}

\bibitem{D3} I. E. Dzyaloshinskii, Theory of helicoidal structures in antiferromagnets.II.metals, 
\href{http://www.jetp.ras.ru/cgi-bin/dn/e_020_01_0223.pdf}{Sov. Phys. JETP \textbf{20}, 223 (1965)}

\bibitem{M2} T. Moriya and T. Miyadai, Evidence for the helical spin structure due to antisymmetric exchange interaction in $Cr_{13}
NbS_2$, \href{https://doi.org/10.1016/0038-1098(82)91006-7}{Solid State Commun. \textbf{42}, 209 (1982).}

\bibitem{miya} T. Miyadai, K. Kikuchi, H. Kondo, S. Sakka, M. Arai, and Y. Ishikawa,Magnetic Properties of $Cr_{1/3}NbS_2$, \href{https://doi.org/10.1143/JPSJ.52.1394}{J. Phys. Soc. Jpn. \textbf{52}, 1394 (1983).}

\bibitem{togawa} Y. Togawa, T. Koyama, K. Takayanagi, S. Mori, Y. Kousaka, J. Akimitsu, S. Nishihara, K. Inoue, A. S. Ovchinnikov, and J. Kishine, Chiral magnetic soliton lattice on a chiral helimagnet, 
\href{https://doi.org/10.1103/PhysRevLett.108.107202}{Phys. Rev. Lett. \textbf{108}, 107202 (2012).}

\bibitem{uk} U. K. Rößler, A. N. Bogdanov, and C. Pfleiderer, Spontaneous skyrmion ground states in magnetic metals, \href{https://doi.org/10.1038/nature05056}{Nature \textbf{442}, 797 (2006).}

\bibitem{muhl} S. Mühlbauer, B. Binz, F. Jonietz, C. Pfleiderer, A. Rosch, A. Neubauer, R. Georgii, and P. Böni, Skyrmion lattice in a chiral magnet, \href{https://doi.org/10.1126/science.1166767}{Science \textbf{323}, 915 (2009).}

\bibitem{yu} X. Z. Yu, Y. Onose, N. Kanazawa, J. H. Park, J. H. Han, Y. Matsui, N. Nagaosa, and Y. Tokura, Real-space observation of a two-dimensional skyrmion crystal, \href{https://doi.org/10.1038/nature09124}{Nature \textbf{465}, 901 (2010).}

\bibitem{heinze} S. Heinze, K. von Bergmann, M. Menzel, J. Brede, A. Kubetzka, R. Wiesendanger, G. Bihlmayer, and S. Blügel, Spontaneous atomic-scale magnetic skyrmion lattice in two dimensions, \href{https://doi.org/10.1038/nphys2045}{Nat. Phys. \textbf{7}, 713 (2011).}

\bibitem{mohapatra1} S. Mohapatra, D. K. Singh, R. Ray, S. Ghosh and A. Singh, Spin-orbit coupling, orbitally entangled antiferromagnetic order, and collective spin-orbital excitations in Sr$_2$VO$_4$, \href{https://iopscience.iop.org/article/10.1088/1361-648X/aca63e}{J. Phys.: Condens. Matter {\bf 35} 045801 (2022).}

\bibitem{mohapatra2} S. Mohapatra, D. K. Singh and A. Singh, Spin-orbit coupling and magnetism in Sr$_2$CrO$_4$, \href{https://iopscience.iop.org/article/10.1088/1361-648X/ace872}{J. Phys.: Condens. Matter {\bf 35} 435601 (2023).}

\bibitem{onose} Y. Onose, T. Ideue, H. Katsura, Y. Shiomi, N. Nagaosa, and Y. Tokura, Observation of the magnon hall effect, \href{https://doi.org/10.1126/science.1188260} {Science \textbf{329}, 297 (2010).}




\bibitem{ryu} K.-S. Ryu, L. Thomas, S.-H. Yang, and S. Parkin, Chiral spin torque at magnetic domain walls, \href{https://doi.org/10.1038/nnano.2013.102}{Nat. Nanotechnol. \textbf{8}, 527 (2013).}

\bibitem{emori} S. Emori, U. Bauer, S.-M. Ahn, E. Martinez, and G. S. D. Beach, Current-driven dynamics of chiral ferromagnetic domain walls, \href{https://doi.org/10.1038/nmat3675}{Nat. Mater. \textbf{12}, 611 (2013).}

\bibitem{jamali} M. Jamali, J. H. Kwon, S.-M. Seo, K.-J. Lee, and H. Yang, Spin wave nonreciprocity for logic device applications, \href{https://doi.org/10.1038/srep03160}{Sci. Rep. \textbf{3}, 3160 (2013).}

\bibitem{garcia} F. Garcia-Sanchez, P. Borys, A. Vansteenkiste, J.-V. Kim, and R. L. Stamps, Nonreciprocal spin-wave channeling along textures driven by the Dzyaloshinskii-Moriya interaction, \href{https://doi.org/10.1103/PhysRevB.89.224408}{Phys. Rev. B \textbf{89}, 224408 (2014)}.

\bibitem{garcia2} F. Garcia-Sanchez, P. Borys, R. Soucaille, J.-P. Adam, R. L. Stamps, and J.-V. Kim, Narrow Magnonic Waveguides Based on Domain Walls, \href{https://doi.org/10.1103/PhysRevLett.114.247206}{Phys. Rev. Lett. \textbf{114}, 247206 (2015).}

\bibitem{lan} J. Lan, W. Yu, and J. Xiao, Antiferromagnetic domain wall as spin wave polarizer and retarder, \href{https://doi.org/10.1038/s41467-017-00265-5}{Nat. Commun. \textbf{8}, 178 (2017).}

\bibitem{cheng} R. Cheng, M. W. Daniels, J.-G. Zhu, and D. Xiao, Antiferromagnetic spin wave field-effect transistor, \href{https://doi.org/10.1038/srep24223}{Sci. Rep. \textbf{6}, 24223 (2016).}

\bibitem{kosty} M. P. Kostyleva, A. A. Serga, T. Schneider, B. Leven, and B. Hillebrands, Spin wave logical gates, \href{https://doi.org/10.1063/1.2089147}{Appl. Phys. Lett. \textbf{87}, 153501 (2005).}



\bibitem{ma} F. Ma and Y. Zhou, Interfacial Dzyaloshinskii–Moriya interaction induced nonreciprocity of spin waves in magnonic waveguides, \href{https://doi.org/10.1039/C4RA07326F}{RSC Adv. \textbf{4}, 46454 (2014).}

\bibitem{lan2} J. Lan, W. Yu, R. Wu, and J. Xiao, Spin-wave diode, \href{https://doi.org/10.1103/PhysRevX.5.041049}{Phys. Rev. X \textbf{5}, 041049 (2015).}

\bibitem{udvardi} L. Udvardi and L. Szunyogh, Chiral asymmetry of the spin-wave spectra in ultrathin magnetic films, \href{https://doi.org/10.1103/PhysRevLett.102.207204}{Phys. Rev. Lett. \textbf{102}, 207204 (2009).}

\bibitem{costa} A. T. Costa, R. B. Muniz, S. Lounis, A. B. Klautau, and D. L. Mills, Spin-orbit coupling and spin waves in ultrathin ferromagnets: The spin-wave Rashba effect, \href{https://doi.org/10.1103/PhysRevB.82.014428}{Phys. Rev. B \textbf{82}, 014428 (2010).}

\bibitem{moon} J.-H. Moon, S.-M. Seo, K.-J. Lee, K.-W. Kim, J. Ryu, H.-W. Lee, R. D. McMichael, and M. D. Stiles, Spin-wave propagation in the presence of interfacial Dzyaloshinskii-Moriya interaction, \href{https://doi.org/10.1103/PhysRevB.88.184404}{Phys. Rev. B \textbf{88}, 184404 (2013).}



\bibitem{santos2} F. José dos Santos, M. dos Santos Dias, and S. Louni, Nonreciprocity of spin waves in noncollinear magnets due to the Dzyaloshinskii-Moriya interaction, \href{https://doi.org/10.1103/PhysRevB.102.104401}{Phys. Rev. B \textbf{102}, 104401 (2020).}


\bibitem{cort} D. Cortés-Ortuño and P. Landeros, Influence of the Dzyaloshinskii-Moriya interaction on the spin-wave spectra of thin films, \href{https://doi.org/10.1088/0953-8984/25/15/156001}{J. Phys.: Condens. Matter \textbf{25}, 156001 (2013).}



\bibitem{markov} I. L. Markov, Limits on fundamental limits to computation, \href{https:/doi.org/10.1038/nature13570}{Nature \textbf{512}, 147 (2014).}

\bibitem{kaji} Y. Kajiwara, K. Harii, S. Takahashi, J. Ohe, K. Uchida1, M. Mizuguchi, H. Umezawa, H. Kawai, K. Ando, and K. Taka \textit{et al.}, Transmission of electrical signals by spin-wave interconversion in a magnetic insulator, \href{https://doi.org/10.1038/nature08876}{Nature \textbf{464}, 262 (2010).}

\bibitem{dieny} B. Dieny, I. L. Prejbeanu, K. Garello , P. Gambardella , P. Freitas, R. Lehndorff, W. Raberg, U. Ebels, S. O. Demokritov, and J. Akerman \textit{et al.}, Opportunities and challenges for spintronics in the microelectronics industry, \href{https://doi.org/10.1038/s41928-020-0461-5}{Nat. Electron. \textbf{3}, 446 (2020).}


\bibitem{bloch} F. Bloch, Zur theorie des ferromagnetismus, \href{https://doi.org/10.1007/BF01339661}{Z. Physik \textbf{61}, 206 (1930).}
  

\bibitem{neusser} S. Neusser and D. Grundler, Magnonics: Spin waves on the nanoscale, \href{https://doi.org/10.1002/adma.200900809}{Adv. Mater. \textbf{21}, 2927 (2009).}


\bibitem{pet}A. N. Petsch, N. S. Headings, D. Prabhakaran, A. I. Kolesnikov, C. D. Frost, A. T. Boothroyd, R. Coldea, and S. M. Hayden., High-energy spin waves in the spin-1 square-lattice antiferromagnet $La_2NiO_4$, \href{https://doi.org/10.1103/PhysRevResearch.5.033113}{Phys. Rev. Research \textbf{5}, 033113 (2023).}

\bibitem{poel} G. Poelchen, J. Hellwig, M. Peters, D. Yu. Usachov, K. Kliemt, C. Laubschat, P. M. Echenique, E. V. Chulkov, C. Krellner, and S. S. P. Parkin \textit{et al.}, Long-lived spin waves in a metallic antiferromagnet, \href{https://doi.org/10.1038/s41467-023-40963-x}{Nat. Commun. 14, 5422 (2023).} 

\bibitem{kim} J. Kim, D. Casa, M. H. Upton, T. Gog, Y. -J. Kim, J. F. Mitchell, M. van Veenendaal, M. Daghofer, J. van den Brink, and G. Khaliullin \textit{et al.}, Magnetic excitation spectra of $Sr_2IrO_4$ probed by resonant inelastic x-ray scattering: Establishing links to cuprate superconductors, \href{http://dx.doi.org/10.1103/PhysRevLett.108.177003}{Phys. Rev. Lett. 108, 177003 (2012).} 




\bibitem{pli} M. Plihal, D. L. Mills, and J. Kirschner, 
Spin wave signature in the spin-polarized electron energy loss spectrum of ultrathin Fe films: Theory and experiment, \href{https://doi.org/10.1103/PhysRevLett.82.2579}{Phys. Rev. Lett. \textbf{82}, 2579 (1999).}

\bibitem{vollmer} R. Vollmer, M. Etzkorn, P. S. A. Kumar, H. Ibach, and J. Kirschner, 
Spin-polarized electron energy loss spectroscopy of high energy, large wave vector spin waves in ultrathin fcc Co films on Cu(001), \href{https://doi.org/10.1103/PhysRevLett.91.147201}{Phys. Rev. Lett. \textbf{91}, 147201 (2003).}

\bibitem{Zakeri} Kh. Zakeri, Y. Zhang, J. Prokop, T.-H. Chuang, N. Sakr, W. X. Tang, and J. Kirschner, Asymmetric spin-wave dispersion on Fe(110): Direct evidence of the Dzyaloshinskii-Moriya interaction, \href{https://doi.org/10.1103/PhysRevLett.104.137203}{Phys. Rev. Lett. \textbf{104}, 137203 (2010).}

\bibitem{zakeri1} Kh. Zakeri, Y. Zhang, T.-H. Chuang, and J. Kirschner, Magnon Lifetimes on the $Fe(110)$ Surface: The Role of Spin-Orbit Coupling, \href{http://dx.doi.org/10.1103/PhysRevLett.108.197205}{Phys. Rev. Lett. {\bf 108}, 197205 (2012).}

\bibitem{tsurk} S. Tsurkan and Kh. Zaker, Giant Dzyaloshinskii-Moriya interaction in epitaxial Co/Fe bilayers with $C_{2v}$ symmetry, \href{https://doi.org/10.1103/PhysRevB.102.060406}{Phys. Rev. B {\bf 102}, 060406(R) (2020).}

\bibitem{kai} K. Di, V. L. Zhang, H. S. Lim, S. C. Ng, M. H. Kuok, X. Qiu, and H. Yang, Asymmetric spin-wave dispersion due to Dzyaloshinskii-Moriya interaction in an ultrathin Pt/CoFeB film, \href{https://doi.org/10.1063/1.4907173}{Appl. Phys. Lett. \textbf{106}, 052403 (2015).}

\bibitem{bel} M. Belmeguenai, J.-P. Adam, Y. Roussigné, S. Eimer, T. Devolder, J.-V. Kim, S. M. Cherif, A. Stashkevich, and A. Thiaville, Interfacial Dzyaloshinskii-Moriya interaction in perpendicularly magnetized $Pt/Co/AlO_x$ ultrathin films measured by Brillouin light spectroscopy, \href{https://doi.org/10.1103/PhysRevB.91.180405}{Phys. Rev. B \textbf{91}, 180405(R) (2015).}




bibitem{ak} A. K. Chaurasiya, C. Banerjee, S. Pan, S. Sahoo, S. Choudhury, J. Sinha, and A. Barman, Direct Observation of Interfacial Dzyaloshinskii-Moriya Interaction from Asymmetric Spin-wave Propagation in $W/CoFeB/SiO_2$ Heterostructures Down to Subnanometer $CoFeB$ Thickness, \href{https://doi.org/10.1038/srep32592}{Sci. Rep. {\bf 6}, 32592 (2016).}

\bibitem{heins} C. Heins, V. Iurchuk, O. Gladii, L. Korber, A. Kakay, J. Fassbender, K. Schultheiss, and H. Schultheiss, Nonreciprocal spin-wave dispersion in magnetic bilayers, \href{ https://doi.org/10.1103/PhysRevB.111.134434}{Phys. Rev. B {\bf 111}, 134434 (2025).}


\bibitem{kostyl} M. Kostylev, Interface boundary conditions for dynamic magnetization and spin wave dynamics in a ferromagnetic layer with the interface Dzyaloshinskii-Moriya interaction, \href{https://doi.org/10.1063/1.4884309}{J. Appl. Phys. \textbf{115}, 233902 (2014).}

\bibitem{kennedy} W. Kennedy, S. dos Anjos Sousa-Júnior, N. C. Costa, and R. R. dos Santos, Magnetism and metal-insulator transitions in the Rashba-Hubbard model, \href{https://doi.org/10.1103/PhysRevB.106.165121}{Phys. Rev. B \textbf{106}, 165121 (2022).}

\bibitem{kubo} K. Kubo, Weyl semimetallic state in the Rashba–Hubbard model, \href{https://doi.org/10.7566/JPSJ.93.024708}{J. Phys. Soc. Jpn. \textbf{93}, 024708 (2024).}

\bibitem{ajain} A. Jain, G. Goyal, and D. K. Singh, Weyl semimetallic state with antiferromagnetic order in the Rashba-Hubbard model, \href{https://doi.org/10.1103/PhysRevB.110.075134}{Phys. Rev. B \textbf{110}, 075134 (2024).}

\bibitem{ggoyal} G. Goyal and D. K. Singh, Antiferromagnetically ordered topological semimetals in Hubbard model with spin-orbit coupling, \href{https://iopscience.iop.org/article/10.1088/1361-648X/ad3792}{J. Phys.: Condens. Matter \textbf{36}, 265802 (2024).}

\bibitem{hotta} M. Kawano and C. Hotta, Phase diagram of the square-lattice Hubbard model with Rashba-type antisymmetric spin-orbit coupling, \href{https://doi.org/10.1103/PhysRevB.107.045123}{Phys. Rev. B \textbf{107}, 045123 (2023).}

\bibitem{sebas} S. dos Anjos Sousa-Júnior and R. Mondaini, Weyl semimetallic, Néel, spiral, and vortex states in the Rashba-Hubbard model, \href{https://doi.org/10.48550/arXiv.2501.01590}{arXiv:2501.01590 (2025).}

\bibitem{sudhakar} S. Pandey and A. Singh, Quantum and thermal fluctuations in a two-dimensional correlated band ferromagnet: Goldstone-mode-preserving investigation with self-energy and vertex corrections, \href{https://doi.org/10.1103/PhysRevB.76.104437}{Phys. Rev. B \textbf{76}, 104437 (2007).}

\bibitem{greco} A. Greco, M. Bejas, and A. P. Schnyder, Ferromagnetic fluctuations in the Rashba-Hubbard model, \href{https://doi.org/10.1103/PhysRevB.101.174420}{Phys. Rev. B \textbf{101}, 174420 (2020).}

\bibitem{sene}D. Sénéchal, D. Perez, and D. Plouffe, Cluster perturbation theory for Hubbard models, \href{https://doi.org/10.1103/PhysRevB.66.075129}{Phys. Rev. B \textbf{66}, 075129 (2002).}


\bibitem{georges} A. Georges, G. Kotliar, W. Krauth, and M. J. Rozenberg, Dynamical mean-field theory of strongly correlated fermion systems and the limit of infinite dimensions, \href{https://doi.org/10.1103/RevModPhys.68.13}{Rev. Mod. Phys. {\bf 68}, 13 (1996).}

\bibitem{vekic} M. Veki\'{c} and S. R. White, Pseudogap formation in the half-filled Hubbard model, \href{https://doi.org/10.1103/PhysRevB.47.1160}{Phys. Rev. B {\bf 47}, 1160 (1993).} 

\bibitem{kubo2} K. Kubo, Variational Monte Carlo study of ferromagnetism in the two-orbital Hubbard model on a square lattice, \href{https://doi.org/10.1103/PhysRevB.79.020407}{Phys. Rev. B {\bf 79}, 020407 (2009).} 


\bibitem{spandey} S. Pandey and A. Singh, Ferromagnetism in the $t-t'$ Hubbard model: Interplay of lattice, band dispersion, and interaction effects studied within a Goldstone-mode preserving scheme, \href{http://dx.doi.org/10.1103/PhysRevB.76.104437}{Phys. Rev. B \textbf{75}, 064412 (2007).}


\bibitem{fark} P. Farkasovsky, Ferromagnetism in the Hubbard model with a generalized type of hopping, \href{https://doi.org/10.1103/PhysRevB.66.012404}{Phys. Rev. B \textbf{66}, 012404 (2002).}

\bibitem{fark2} P. Farkasovsky, Ferromagnetism in the asymmetric Hubbard model, \href{http://dx.doi.org/10.1140/epjb/e2012-30306-9}{Eur. Phys. J. B \textbf{85}, 253 (2012).}

\bibitem{fark3} P. Farkasovsky, Ferromagnetism in the Hubbard Model with Long-range and Correlated Hopping, \href{https://doi.org/10.1007/s10582-004-0110-7}{Czech. J. Phys. \textbf{54}, 419 (2004).} 

\bibitem{abate} E. Abate and M. Asdente, Tight-Binding Calculation of 3$d$ Bands of Fe with and without Spin-Orbit Coupling, \href{https://doi.org/10.1103/PhysRev.140.A1303}{Phys. Rev. {\bf 140}, 1303 (1965).}

\bibitem{zakeri2} K. Zakeri and A. von Faber, Giant Spin-Orbit Induced Magnon Nonreciprocity in Ultrathin Ferromagnets, \href{https://doi.org/10.1103/PhysRevLett.132.126702}{Phys. Rev. Lett. {\bf 132}, 126702 (2024).}


\end{thebibliography}
\end{document}